\documentclass[12pt]{article}
 \usepackage{epsfig}
 \usepackage{pstricks}
 \usepackage{pst-coil}

\def\sbs#1{{}_{\mbox{\scriptsize #1}}}          
\newcommand{\gl}{\tilde{g}}                     
\newcommand{\wi}{\tilde{W_3}}                   
\newcommand{\bi}{\tilde{b}}                     
\newcommand{\hi}{\tilde{h_1^0}}                 
\newcommand{\hj}{\tilde{h_2^0}}                 
\newcommand{\gr}{\tilde{G}}

\newcommand{\mmoo}{{(m_D^{\dg}m_D)_{11}}}

\newcommand{\snone}{\widetilde{N_1^c}}

\def\c{\chi}

\def\e{\epsilon}
\def\f{\phi}
\def\g{\gamma}

\def\j{\psi}
\def\k{\kappa}
\def\l{\lambda}
\def\m{\mu}
\def\n{\nu}

\def\p{\pi}

\def\r{\rho}
\def\s{\sigma}
\def\t{\tau}

\def\z{\zeta}
\def\D{\Delta}

\def\G{\Gamma}

\def\O{\Omega}

\def\ve{\varepsilon}
\def\vf{\varphi}

\def\cl{{\cal L}}

\def\co{{\cal O}}

\def\bo{{\raise.15ex\hbox{\large$\Box$}}}               
\def\pr{\prod}                                          
\def\ltap{\raisebox{-.4ex}{\rlap{$\sim$}} \raisebox{.4ex}{$<$}}   
\def\face{{\raise.2ex\hbox{$\displaystyle \bigodot$}\mskip-2.2mu \llap {$\ddot
        \smile$}}}                                      
\def\dg{\dagger}                                     

\def\wt#1{\widetilde{#1}}                    
\def\VEV#1{\left\langle #1\right\rangle}        
\def\leftrightarrowfill{$\mathsurround=0pt \mathord\leftarrow \mkern-6mu
        \cleaders\hbox{$\mkern-2mu \mathord- \mkern-2mu$}\hfill
        \mkern-6mu \mathord\rightarrow$}       
\def\dvec#1{\vbox{\ialign{##\crcr
        \leftrightarrowfill\crcr\noalign{\kern-1pt\nointerlineskip}
        $\hfil\displaystyle{#1}\hfil$\crcr}}}           

\def\beq{\begin{equation}}
\def\eeq{\end{equation}}

\def\beqx{\begin{displaymath}}
\def\eeqx{\end{displaymath}}

\def\beqa{\begin{eqnarray}}
\def\eeqa{\end{eqnarray}}
\def\NO{\nonumber}

\def\pl#1#2#3{Phys.~Lett.~{\bf B {#1}} (19{#2}) #3}
\def\np#1#2#3{Nucl.~Phys.~{\bf B {#1}} (19{#2}) #3}
\def\prl#1#2#3{Phys.~Rev.~Lett.~{\bf #1} (19{#2}) #3}
\def\pr#1#2#3{Phys.~Rev.~{\bf D {#1}} (19{#2}) #3}

\def\prep#1#2#3{Phys.~Rep.~{\bf {#1}C} (19{#2}) #3}
\def\ptp#1#2#3{Progr.~Theor.~Phys.~{\bf {#1}} (19{#2}) #3}

\def\nc#1#2#3{Nuovo Cim.~{\bf {#1}} (19{#2}) #3}

\catcode`@=11
\def\@citex[#1]#2{\if@filesw\immediate\write\@auxout{\string\citation{#2}}\fi
  \def\@citea{}\@cite{\@for\@citeb:=#2\do
    {\@citea\def\@citea{,\penalty\@m}\@ifundefined
      {b@\@citeb}{{\bf ?}\@warning
       {Citation `\@citeb' on page \thepage \space undefined}}%
\hbox{\csname b@\@citeb\endcsname}}}{#1}}

\def\citer{\@ifnextchar [{\@tempswatrue\@citexr}{\@tempswafalse\@citexr[]}}
\def\@citexr[#1]#2{\scriptsize 
  \if@filesw\immediate\write\@auxout{\string\citation{#2}}\fi
  \def\@citea{}\@cite{\@for\@citeb:=#2\do
    {\@citea\def\@citea{-\penalty\@m}\@ifundefined
       {b@\@citeb}{{\bf ?}\@warning
       {Citation `\@citeb' on page \thepage \space undefined}}%
\hbox{\csname b@\@citeb\endcsname}}}{#1}\normalsize}
\catcode`@=12

\catcode`\@=11
\long\def\@makefntext#1{ 
\protect\noindent \hbox to 3.2pt {\hskip-.9pt  
$^{{\ninerm\@thefnmark}}$\hfil}#1\hfill} 

\def\thefootnote{\fnsymbol{footnote}}
 \def\@makefnmark{\hbox to 0pt{$^{\@thefnmark}$\hss}}  
        
\def\ps@myheadings{\let\@mkboth\@gobbletwo
\def\@oddhead{\hbox{} 
\rightmark\hfil\ninerm\thepage}   
\def\@oddfoot{}\def\@evenhead{\ninerm\thepage\hfil 
\leftmark\hbox{}}\def\@evenfoot{}
\def\sectionmark##1{}\def\subsectionmark##1{}}

\begin{document}

\newcommand{\symbolfootnote}{\renewcommand{\thefootnote}
        {\fnsymbol{footnote}}}
\renewcommand{\thefootnote}{\fnsymbol{footnote}}
\newcommand{\alphfootnote}
        {\setcounter{footnote}{0}
         \renewcommand{\thefootnote}{\sevenrm\alph{footnote}}}

\newcounter{sectionc}\newcounter{subsectionc}\newcounter{subsubsectionc}
\renewcommand{\section}[1] {\vspace{0.6cm}\addtocounter{sectionc}{1} 
\setcounter{subsectionc}{0}\setcounter{subsubsectionc}{0}\noindent 
        {\bf\thesectionc. #1}\par\vspace{0.4cm}}
\renewcommand{\subsection}[1] {\vspace{0.6cm}\addtocounter{subsectionc}{1} 
        \setcounter{subsubsectionc}{0}\noindent 
        {\it\thesectionc.\thesubsectionc. #1}\par\vspace{0.4cm}}
\renewcommand{\subsubsection}[1] {\vspace{0.6cm}\addtocounter{subsubsectionc}{1}
        \noindent {\rm\thesectionc.\thesubsectionc.\thesubsubsectionc. 
        #1}\par\vspace{0.4cm}}
\newcommand{\nonumsection}[1] {\vspace{0.6cm}\noindent{\bf #1}
        \par\vspace{0.4cm}}
                                                 
\newcounter{appendixc}
\newcounter{subappendixc}[appendixc]
\newcounter{subsubappendixc}[subappendixc]
\renewcommand{\thesubappendixc}{\Alph{appendixc}.\arabic{subappendixc}}
\renewcommand{\thesubsubappendixc}
        {\Alph{appendixc}.\arabic{subappendixc}.\arabic{subsubappendixc}}

\renewcommand{\appendix}[1] {\vspace{0.6cm}
        \refstepcounter{appendixc}
        \setcounter{figure}{0}
        \setcounter{table}{0}
        \setcounter{equation}{0}
        \renewcommand{\thefigure}{\Alph{appendixc}.\arabic{figure}}
        \renewcommand{\thetable}{\Alph{appendixc}.\arabic{table}}
        \renewcommand{\theappendixc}{\Alph{appendixc}}
        \renewcommand{\theequation}{\Alph{appendixc}.\arabic{equation}}
        \noindent{\bf Appendix \theappendixc #1}\par\vspace{0.4cm}}
\newcommand{\subappendix}[1] {\vspace{0.6cm}
        \refstepcounter{subappendixc}
        \noindent{\bf Appendix \thesubappendixc. #1}\par\vspace{0.4cm}}
\newcommand{\subsubappendix}[1] {\vspace{0.6cm}
        \refstepcounter{subsubappendixc}
        \noindent{\it Appendix \thesubsubappendixc. #1}
        \par\vspace{0.4cm}}

\newcommand{\bibit}{\it}
\newcommand{\bibbf}{\bf}
\renewenvironment{thebibliography}[1]
        {\begin{list}{\arabic{enumi}.}
        {\usecounter{enumi}\setlength{\parsep}{0pt} 
\setlength{\leftmargin 1.25cm}{\rightmargin 0pt} 
         \setlength{\itemsep}{0pt} \settowidth
        {\labelwidth}{#1.}\sloppy}}{\end{list}}

\topsep=0in\parsep=0in\itemsep=0in
\parindent=1.5pc

\newcounter{itemlistc}
\newcounter{romanlistc}
\newcounter{alphlistc}
\newcounter{arabiclistc}
\newenvironment{itemlist}
        {\setcounter{itemlistc}{0}
         \begin{list}{$\bullet$}
        {\usecounter{itemlistc}
         \setlength{\parsep}{0pt}
         \setlength{\itemsep}{0pt}}}{\end{list}}

\newenvironment{romanlist}
        {\setcounter{romanlistc}{0}
         \begin{list}{$($\roman{romanlistc}$)$}
        {\usecounter{romanlistc}
         \setlength{\parsep}{0pt}
         \setlength{\itemsep}{0pt}}}{\end{list}}

\newenvironment{alphlist}
        {\setcounter{alphlistc}{0}
         \begin{list}{$($\alph{alphlistc}$)$}
        {\usecounter{alphlistc}
         \setlength{\parsep}{0pt}
         \setlength{\itemsep}{0pt}}}{\end{list}}

\newenvironment{arabiclist}
        {\setcounter{arabiclistc}{0}
         \begin{list}{\arabic{arabiclistc}}
        {\usecounter{arabiclistc}
         \setlength{\parsep}{0pt}
         \setlength{\itemsep}{0pt}}}{\end{list}}

\newcommand{\fcaption}[1]{
        \refstepcounter{figure}
        \setbox\@tempboxa = \hbox{\tenrm Fig.~\thefigure. #1}
        \ifdim \wd\@tempboxa > 6in
           {\begin{center}
        \parbox{6in}{\tenrm\baselineskip=12pt Fig.~\thefigure. #1 }
            \end{center}}
        \else
             {\begin{center}
             {\tenrm Fig.~\thefigure. #1}
              \end{center}}
        \fi}

\newcommand{\tcaption}[1]{
        \refstepcounter{table}
        \setbox\@tempboxa = \hbox{\tenrm Table~\thetable. #1}
        \ifdim \wd\@tempboxa > 6in
           {\begin{center}
        \parbox{6in}{\tenrm\baselineskip=12pt Table~\thetable. #1 }
            \end{center}}
        \else
             {\begin{center}
             {\tenrm Table~\thetable. #1}
              \end{center}}
        \fi}

\def\@citex[#1]#2{\if@filesw\immediate\write\@auxout
        {\string\citation{#2}}\fi
\def\@citea{}\@cite{\@for\@citeb:=#2\do
        {\@citea\def\@citea{,}\@ifundefined
        {b@\@citeb}{{\bf ?}\@warning
        {Citation `\@citeb' on page \thepage \space undefined}}
        {\csname b@\@citeb\endcsname}}}{#1}}

\newif\if@cghi
\def\cite{\@cghitrue\@ifnextchar [{\@tempswatrue
        \@citex}{\@tempswafalse\@citex[]}}
\def\citelow{\@cghifalse\@ifnextchar [{\@tempswatrue
        \@citex}{\@tempswafalse\@citex[]}}
\def\@cite#1#2{{$\null^{#1}$\if@tempswa\typeout
        {IJCGA warning: optional citation argument 
        ignored: `#2'} \fi}}
\newcommand{\citeup}{\cite}

\def\fnm#1{$^{\mbox{\scriptsize #1}}$}
\def\fnt#1#2{\footnotetext{\kern-.3em
        {$^{\mbox{\sevenrm #1}}$}{#2}}}

\font\twelvebf=cmbx10 scaled\magstep 1
\font\twelverm=cmr10 scaled\magstep 1
\font\twelveit=cmti10 scaled\magstep 1
\font\elevenbfit=cmbxti10 scaled\magstephalf
\font\elevenbf=cmbx10 scaled\magstephalf
\font\elevenrm=cmr10 scaled\magstephalf
\font\elevenit=cmti10 scaled\magstephalf
\font\bfit=cmbxti10
\font\tenbf=cmbx10
\font\tenrm=cmr10
\font\tenit=cmti10
\font\ninebf=cmbx9
\font\ninerm=cmr9
\font\nineit=cmti9
\font\eightbf=cmbx8
\font\eightrm=cmr8
\font\eightit=cmti8

\date{}
\title{
{\large\rm DESY 98-171}\hfill{\mbox{}}\\
{\large\rm December 1998}\hfill{\mbox{}}\vspace*{3cm}\\
{\bf BARYOGENESIS ABOVE\\ THE FERMI SCALE}
\thanks{presented at the 5th Colloque Cosmologie, Paris, June 1998}}
\author{W. Buchm\"uller \\
\vspace{3.0\baselineskip}                                               
{\normalsize\it Deutsches Elektronen-Synchrotron DESY, 22603 Hamburg, Germany}
\vspace*{2cm}\\                     
}        

\maketitle

\thispagestyle{empty}

\begin{abstract}
\noindent
In the standard model and most of its extensions the electroweak transition
is too weak to affect the cosmological baryon asymmetry. Due to sphaleron
processes baryogenesis in the high-temperature, symmetric phase of the 
standard model is closely related to neutrino properties. The experimental
indications for very small neutrino masses from the solar and the 
atmospheric neutrino deficits favour a large scale of $B-L$ breaking.
For hierarchical neutrino masses, with $B-L$ broken at the
unification scale $\Lambda_{\mbox{\scriptsize GUT}}\sim 10^{16}\;$GeV,  
the observed baryon asymmetry $n_B/s \sim 10^{-10}$ is naturally
explained by the decay of heavy Majorana neutrinos. The corresponding
baryogenesis temperature is $T_B \sim 10^{10}$ GeV. In supersymmetric
models implications for the mass spectrum of superparticles can be
derived. A consistent picture is obtained with the gravitino as LSP,
which may be the dominant component of cold dark matter.
\end{abstract}

\newpage

\section{General aspects of baryogenesis}

The cosmological baryon asymmetry, the ratio of the baryon density to the
entropy density of the universe,
\beq
Y_B = {n_B\over s} = (0.6 - 1)\cdot 10^{-10}\;,
\eeq
can be understood as a consequence of baryon number violation, C and
CP violation, and a deviation from thermal equilibrium\cite{sac}. The
presently observed value of the baryon asymmetry is then explained as a 
consequence of the spectrum and interactions of elementary particles, 
together with the cosmological evolution.

\begin{figure}[t]
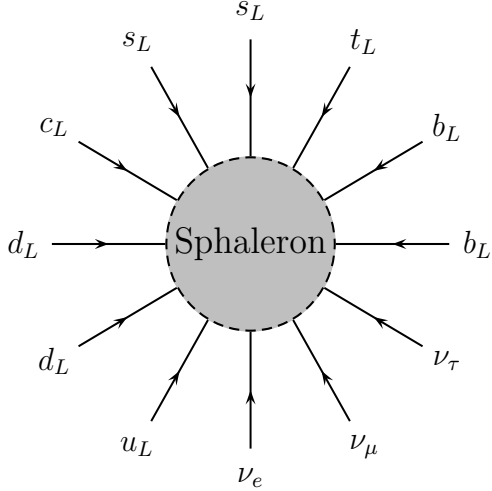

    \begin{center}
      \scaleboxto(7,7){
        \parbox[c]{9cm}{ \begin{center}
     \pspicture*(-0.50,-2.5)(8.5,6.5)
     \psset{linecolor=lightgray}
     \qdisk(4,2){1.5cm}
     \psset{linecolor=black}
     \pscircle[linewidth=1pt,linestyle=dashed](4,2){1.5cm}
     \rput[cc]{0}(4,2){\scalebox{1.5}{Sphaleron}}
     \psline[linewidth=1pt](5.50,2.00)(7.50,2.00)
     \psline[linewidth=1pt](5.30,2.75)(7.03,3.75)
     \psline[linewidth=1pt](4.75,3.30)(5.75,5.03)
     \psline[linewidth=1pt](4.00,3.50)(4.00,5.50)
     \psline[linewidth=1pt](3.25,3.30)(2.25,5.03)
     \psline[linewidth=1pt](2.70,2.75)(0.97,3.75)
     \psline[linewidth=1pt](2.50,2.00)(0.50,2.00)
     \psline[linewidth=1pt](2.70,1.25)(0.97,0.25)
     \psline[linewidth=1pt](3.25,0.70)(2.25,-1.03)
     \psline[linewidth=1pt](4.00,0.50)(4.00,-1.50)
     \psline[linewidth=1pt](4.75,0.70)(5.75,-1.03)
     \psline[linewidth=1pt](5.30,1.25)(7.03,0.25)
     \psline[linewidth=1pt]{<-}(6.50,2.00)(6.60,2.00)
     \psline[linewidth=1pt]{<-}(6.17,3.25)(6.25,3.30)
     \psline[linewidth=1pt]{<-}(5.25,4.17)(5.30,4.25)
     \psline[linewidth=1pt]{<-}(4.00,4.50)(4.00,4.60)
     \psline[linewidth=1pt]{<-}(2.75,4.17)(2.70,4.25)
     \psline[linewidth=1pt]{<-}(1.83,3.25)(1.75,3.30)
     \psline[linewidth=1pt]{<-}(1.50,2.00)(1.40,2.00)
     \psline[linewidth=1pt]{<-}(1.83,0.75)(1.75,0.70)
     \psline[linewidth=1pt]{<-}(2.75,-0.17)(2.70,-0.25)
     \psline[linewidth=1pt]{<-}(4.00,-0.50)(4.00,-0.60)
     \psline[linewidth=1pt]{<-}(5.25,-0.17)(5.30,-0.25)
     \psline[linewidth=1pt]{<-}(6.17,0.75)(6.25,0.70)
     \rput[cc]{0}(8.00,2.00){\scalebox{1.3}{$b_L$}}
     \rput[cc]{0}(7.46,4.00){\scalebox{1.3}{$b_L$}}
     \rput[cc]{0}(6.00,5.46){\scalebox{1.3}{$t_L$}}
     \rput[cc]{0}(4.00,6.00){\scalebox{1.3}{$s_L$}}
     \rput[cc]{0}(2.00,5.46){\scalebox{1.3}{$s_L$}}
     \rput[cc]{0}(0.54,4.00){\scalebox{1.3}{$c_L$}}
     \rput[cc]{0}(0.00,2.00){\scalebox{1.3}{$d_L$}}
     \rput[cc]{0}(0.54,0.00){\scalebox{1.3}{$d_L$}}
     \rput[cc]{0}(2.00,-1.46){\scalebox{1.3}{$u_L$}}
     \rput[cc]{0}(4.00,-2.00){\scalebox{1.3}{$\nu_e$}}
     \rput[cc]{0}(6.00,-1.46){\scalebox{1.3}{$\nu_{\mu}$}}
     \rput[cc]{0}(7.46,0.00){\scalebox{1.3}{$\nu_{\tau}$}}
     \endpspicture
\end{center}}
      }
    \end{center}
    \caption{\it One of the 12-fermion processes which are in thermal 
      equilibrium in the high-temperature phase of the standard model.
      \label{fig_sphal}
    }
  \end{figure}

A crucial ingredient of baryogenesis is the connection between baryon number
($B$) and lepton number ($L$) in the high-temperature, symmetric phase of
the standard model. Due to the chiral nature of the weak interactions $B$ and
$L$ are not conserved\cite{thoo}. At zero 
temperature this has no observable effect due to the smallness of the weak 
coupling. However, as the temperature approaches the critical temperature 
$T_c$ of the electroweak phase transition, $B$ and $L$ violating processes 
come into thermal equilibrium\cite{krs}. The rate of these processes is
related to the free energy of sphaleron-type field configurations which carry
topological charge. In the standard model they lead to an effective
interaction of all left-handed fermions (cf. fig.~\ref{fig_sphal}) which 
violates baryon and lepton number by three units, 
\beq 
    \D B = \D L = 3\;. \label{sphal1}
\eeq
The evaluation of the sphaleron rate in the symmetric high temperature phase is
a challenging problem\cite{arn}. Although a complete theoretical understanding
has not yet been achieved, it is generally believed that $B$ and $L$ violating
processes are in thermal equilibrium for temperatures in the range
\beq 
T_c^{EW} \sim 100\ \mbox{GeV} < T < T_{SPH} \sim 10^{12}\ \mbox{GeV}\;.
\eeq

Sphaleron processes have a profound effect on the generation of the
cosmological baryon asymmetry.  Eq.~\ref{sphal1} suggests that any
$B+L$ asymmetry generated before the electroweak phase transition,
i.e., at temperatures $T>T_c^{EW}$, will be washed out. However, since
only left-handed fields couple to sphalerons, a non-zero value of
$B+L$ can persist in the high-temperature, symmetric phase if there
exists a non-vanishing $B-L$ asymmetry. An analysis of the chemical potentials
of all particle species in the high-temperature phase yields the following
relation between the baryon asymmetry
$Y_B = (n_B-n_{\overline{B}})/s$ and the corresponding
$L$ and $B-L$ asymmetries $Y_L$ and $Y_{B-L}$, respectively\cite{chem},
\beq\label{basic}
Y_B\ =\ C\ Y_{B-L}\ =\ {C\over C-1}\ Y_L\;.
\eeq
Here $C$ is a number ${\cal O}(1)$. In the standard model with three 
generations and one Higgs doublet one has $C=28/79$. 
  
We conclude that $B-L$ violation is needed if the baryon asymmetry is
generated before the electroweak transition, i.e. at temperatures 
$T > T_c^{EW} \sim 100$~GeV.
In the standard model, as well as its supersymmetric version and its unified 
extensions based on the gauge group SU(5), $B-L$ is a conserved quantity. 
Hence, no baryon asymmetry can be generated dynamically in these models.

  \begin{figure}[b]
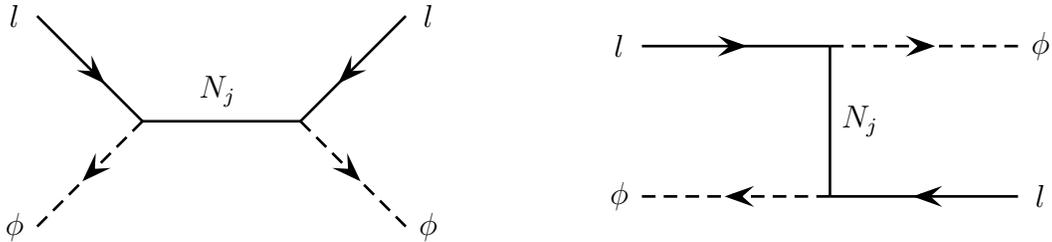

     \begin{center}
  \pspicture[0.5](1,0.3)(8,3.5)
    \psline[linewidth=1pt,linestyle=dashed](1.6,0.6)(3,2)
    \psline[linewidth=2pt]{<-}(2.2,1.2)(2.3,1.3)
    \rput[cc]{0}(1.3,0.6){$\displaystyle \f$}
    \psline[linewidth=1pt](1.6,3.4)(3,2)
    \psline[linewidth=2pt]{->}(2.4,2.6)(2.5,2.5)
    \rput[cc]{0}(1.3,3.4){$\displaystyle l$}
    \psline[linewidth=1pt](3,2)(5.1,2)
    \rput[cc]{0}(4,2.4){$\displaystyle N_j$}
    \psline[linewidth=1pt](5.1,2)(6.5,3.4)
    \psline[linewidth=2pt]{->}(5.7,2.6)(5.6,2.5)
    \rput[cc]{0}(6.8,3.4){$\displaystyle l$}
    \psline[linewidth=1pt,linestyle=dashed](5.1,2)(6.5,0.6)
    \psline[linewidth=2pt]{->}(5.8,1.3)(5.9,1.2)
    \rput[cc]{0}(6.8,0.6){$\displaystyle \f$}
    \endpspicture
    \hspace{1cm}
    \pspicture[0.5](8.5,0.3)(15,3.5)
    \psline[linewidth=1pt](9,3)(11.5,3)
    \psline[linewidth=2pt]{->}(10.3,3)(10.4,3)
    \rput[cc]{0}(8.7,3){$\displaystyle l$}
    \psline[linewidth=1pt,linestyle=dashed](11.5,3)(14,3)
    \psline[linewidth=2pt]{->}(12.8,3)(12.9,3)
    \rput[cc]{0}(14.3,3){$\displaystyle \f$}
    \psline[linewidth=1pt](11.5,3)(11.5,1)
    \rput[cc]{0}(11.9,2){$\displaystyle N_j$}
    \psline[linewidth=1pt,linestyle=dashed](9,1)(11.5,1)
    \psline[linewidth=2pt]{<-}(10.1,1)(10.2,1)
    \rput[cc]{0}(8.7,1){$\displaystyle \f$}
    \psline[linewidth=1pt](11.5,1)(14,1)
    \psline[linewidth=2pt]{<-}(12.6,1)(12.7,1)
    \rput[cc]{0}(14.3,1){$\displaystyle l$}
  \endpspicture
\end{center}
     \caption{\it Lepton number violating lepton Higgs scattering
       \label{lept_fig}}
  \end{figure}

The remnant of lepton number violation in extensions of the standard model
at low energies is the appearance of an effective $\Delta L=2$ interaction 
between lepton and Higgs fields,
  \beq\label{dl2}
  \cl_{\Delta L=2} = {1\over 2}\overline{l_L}\,\f\,g_{\n}\,{1\over M}\,
         g_{\n}^T\,\f\,l_L^c +\mbox{ h.c.}\;.\label{intl2}
  \eeq
Such an interaction arises in particular from the exchange of heavy Majorana
neutrinos (cf.~fig.~\ref{lept_fig}). In the Higgs phase of the standard
model, where the Higgs field aquires a vacuum expectation value, it gives
rise to Majorana masses of the light neutrinos $\n_e$, $\n_\m$ and $\n_\t$.   

At finite temperature the $\Delta L=2$ processes described by (\ref{dl2}) take
place with the rate\cite{fy1}
  \beq
    \Gamma_{\Delta L=2} (T) = {1\over \pi^3}\,{T^3\over v^4}\, 
    \sum_{i=e,\m,\t} m_{\n_i}^2\; .
  \eeq
In thermal equilibrium this yields an additional relation between the
chemical potentials which implies
\beq
Y_B\ =\ Y_{B-L}\ =\ Y_L\ =\ 0 \; .
\eeq

To avoid this conclusion, the $\Delta L=2$ interaction (\ref{intl2}) must not 
reach thermal equilibrium. For baryogenesis at very high temperatures, 
$T > T_{SPH} \sim 10^{12}$ GeV, one has to require 
$\G_{\D L=2} < H|_{T_{SPH}}$,
where $H$ is the Hubble parameter. This yields a stringent upper bound on
Majorana neutrino masses,
\beq\label{nbound}
\sum_{i=e,\m,\t} m_{\n_i}^2 < (0.2\ \mbox{eV})^2\;.
\eeq
This bound is comparable to the upper bound on the electron neutrino mass 
obtained from neutrinoless double beta decay. Note, however, that the bound
also applies to the $\t$-neutrino mass. In supersymmetric
theories two chiral U(1) symmetries in addition to baryon and lepton number
are approximately conserved at temperatures above 
$T_{SS} \sim 10^7$ GeV\cite{iba}. This relaxes the upper bound (\ref{nbound})
from 0.2~eV to about 60~eV.
  
The connection between lepton number and the baryon asymmetry is lost
if baryogenesis takes place at or below the Fermi scale\cite{dol}. However, 
detailed studies of the thermodynamics of the electroweak transition have
shown that, at least in the standard model, the deviation from thermal
equilibrium is not sufficient for baryogenesis\cite{jansen}. In the minimal 
supersymmetric extension of the standard model (MSSM) such a scenario 
appears still possible for a limited range of parameters\cite{dol}.

\section{Decays of heavy Majorana neutrinos}

Baryogenesis above the Fermi scale requires $B-L$ violation, and therefore 
$L$ violation. Lepton number violation is most simply realized by adding 
right-handed Majorana neutrinos to the standard model.  Heavy right-handed 
Majorana neutrinos, whose existence is predicted by theories based on gauge 
groups containing the Pati-Salam symmetry\cite{pat} 
SU(4)$\otimes$SU(2)$_L\otimes$SU(2)$_R$, can also explain the smallness of 
the light neutrino masses via the see-saw mechanism\cite{seesaw}.

The most general lagrangian for couplings and masses of charged
leptons and neutrinos reads 
  \beq\label{yuk}
    \cl_Y = -\overline{l_L}\,\wt{\f}\,g_l\,e_R
            -\overline{l_L}\,\f\,g_{\n}\,\n_R
            -{1\over2}\,\overline{\n^C_R}\,M\,\n_R
            +\mbox{ h.c.}\;.
  \eeq
The vacuum expectation value of the Higgs field $\VEV{\vf^0}=v\ne0$
generates Dirac masses $m_l$ and $m_D$ for charged leptons and neutrinos,
$m_l=g_lv$ and $m_D=g_{\n}v$, respectively, which are assumed to be much 
smaller than the Majorana masses $M$.
This yields light and heavy neutrino mass eigenstates
  \beq
     \n\simeq K^{\dg}\n_L+\n_L^C K\quad,\qquad
     N\simeq\n_R+\n_R^C\, ,
  \eeq
with masses
  \beq
     m_{\n}\simeq- K^{\dg}m_D{1\over M}m_D^T K^*\,
     \quad,\quad  m_N\simeq M\, .
     \label{seesaw}
  \eeq
  Here $K$ is a unitary matrix which relates weak and mass eigenstates. 
  
  The right-handed neutrinos, whose exchange may erase any lepton
  asymmetry, can also generate a lepton asymmetry by means of
  out-of-equilibrium decays. This lepton asymmetry is then partially 
  transformed into a baryon asymmetry by sphaleron processes\cite{fy}.  
  The decay width of the heavy neutrino $N_i$ reads at tree level,
  \beqa
    \G_{Di}&=&\G\left(N^i\to\f^c+l\right)+\G\left(N^i\to\f+l^c\right)\NO\\
           &=&{1\over8\p}{(m_D^{\dg}m_D)_{ii}\over v^2}M_i\;.
    \label{decay}
  \eeqa
From the decay width one obtains an upper bound on the light neutrino masses
  via the out-of-equilibrium condition\cite{fisch}.
  Requiring $\Gamma_{D1}< H|_{T=M_1}$ yields the constraint
  \beq\label{ooeb}
  \wt{m}_1\ =\ {\mmoo\over M_1}\ < \ 10^{-3}\, \mbox{eV}\;.
  \eeq
More direct bounds on the light neutrino masses depend on the structure
of the Dirac neutrino mass matrix as we shall discuss below.

Interference between the tree-level amplitude and the one-loop 
self-energy and vertex corrections yields the $CP$ asymmetry\cite{cov,bp2}
\beqa
\ve_1&=&{\Gamma(N_1\rightarrow l \f^c)-\Gamma(N_1\rightarrow l^c \f)\over
         \Gamma(N_1\rightarrow l \f^c)+\Gamma(N_1\rightarrow l^c \f)}\NO\\
     &\simeq&{3\over16\pi v^2}\;{1\over\left(m_D^{\dag}m_D\right)_{11}}
      \sum_{i=2,3}\mbox{Im}\left[\left(m_D^{\dag}m_D\right)_{1i}^2\right]
      {M_1\over M_i}\label{cpa}\;.
\eeqa
Here we have assumed $M_1\ll M_2,M_3$. For very small mass differences,
which are comparable to the decay widths, one obtains a resonance 
enhancement\cite{pil}. 

The CP asymmetry (\ref{cpa}) leads to the generated lepton 
asymmetry\cite{kw},
\beq
Y_L\ =\ {n_L-n_{\overline{L}}\over s}\ =\ \k\ {\ve_1\over g_*}\;.
    \label{esti}
\eeq
Here the factor $\k<1$ represents the effect of washout processes. In order
to determine $\k$ one has to solve the full Boltzmann equations. In the
examples discussed below one has $\k\simeq 0.1\ldots 0.01$.

The CP asymmetry (\ref{cpa}) is given in terms of the Dirac and the Majorana
neutrino mass matrices. One can always choose a basis for the right-handed
neutrinos where the Majorana mass $M$ is diagonal with real and positive 
eigenvalues. $m_D$ is a general complex matrix, which can be diagonalized by 
a biunitary transformation. One then has
\beq
    m_D=V\,\left(\begin{array}{ccc}m_1&0&0\\0&m_2&0\\0&0&m_3
              \end{array}\right)\,U^{\dag} \;,\quad
    M=\left(\begin{array}{ccc}M_1&0&0\\0&M_2&0\\0&0&M_3
    \end{array}\right)\;,
\eeq
  where $V$ and $U$ are unitary matrices and the $m_i$ are real and
  positive. In the absence of a Majorana mass term $V$ and $U$ would 
  correspond to Kobayashi-Maskawa type mixing matrices of left- and 
  right-handed charged currents, respectively.

 \begin{figure}[t]
    \mbox{ }\hfill
    \epsfig{file=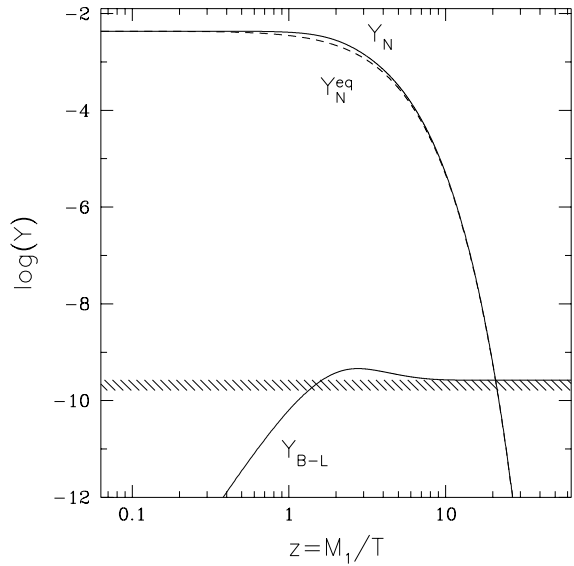,width=8.2cm}
    \hfill\mbox{ }
    \caption{\it Time evolution of the neutrino number density and the
     B-L asymmetry for $\l\simeq 0.1$ and $m_3\simeq m_t$\label{asyNS}}
  \end{figure} 
  
  Note, that according to eqs.~(\ref{decay}) and (\ref{cpa}) the $CP$
  asymmetry is determined by the mixings and phases present in the
  product $m_D^{\dg}m_D$, where the matrix $V$ drops out.  Hence, to
  leading order, the mixings and phases which are responsible for
  baryogenesis are entirely determined by the matrix $U$.
  Correspondingly, the mixing matrix $K$ in the leptonic charged
  current, which determines $CP$ violation and mixings of the light
  leptons, depends on mass ratios and mixing angles and phases of $U$
  and $V$.  This implies that there exists no direct connection
  between $CP$ violation and generation mixing which are relevant at high
  energies and at low energies, respectively.
 
In many models the quark and lepton mass hierarchies and mixings are 
parametrised in terms of a common mixing parameter $\l \sim 0.1$. Assuming
a hierarchy for the right-handed neutrino masses similar to the one
satisfied by up-type quarks,
\beq
{M_1\over M_2}\ \sim\ {M_2\over M_3}\ \sim\ \l^2\;,
\eeq
and a corresponding CP asymmetry 
\beq
\ve_1\ \sim\ {\l^4\over 16\pi}\ {m_3^2\over v^2}\ 
       \sim\ 10^{-6}\ {m_3^2\over v^2} \;,
\eeq 
one obtains indeed the correct order of magnitude for the baryon 
asymmetry\cite{bp} if one chooses $m_3\simeq m_t\simeq 174$ GeV, as expected 
in theories with Pati-Salam symmetry. Using as a constraint the 
value for the $\n_{\m}$-mass which is preferred by the MSW 
explanation\cite{msw} of the solar neutrino deficit\cite{tot}, 
$m_{\n_{\m}}\simeq 3\cdot10^{-3}$~eV, the ansatz\cite{bp} implies for the 
other light and the heavy neutrino masses 
\beq
m_{\n_e}\simeq 8\cdot 10^{-6}\ \mbox{eV}\;,
\quad m_{\n_{\t}}\simeq 0.15\ \mbox{eV}\;, 
\qquad M_3 \simeq 2\cdot10^{14}\ \mbox{GeV}\;. \label{nmass} 
\eeq
Consequently, one has $M_1\simeq 2\cdot10^{10}$ GeV and
$M_2\simeq 2\cdot10^{12}$ GeV. The solution of the Boltzmann equations 
then yields the baryon asymmetry (see fig.~\ref{asyNS}),
  \beq
     Y_B \simeq 9\cdot10^{-11}\; , \label{nonsusy_res1}
  \eeq
which is indeed the correct order of magnitude. The precise value
depends on unknown phases.

  \begin{figure}[t]
     \begin{center}
     \epsfig{file=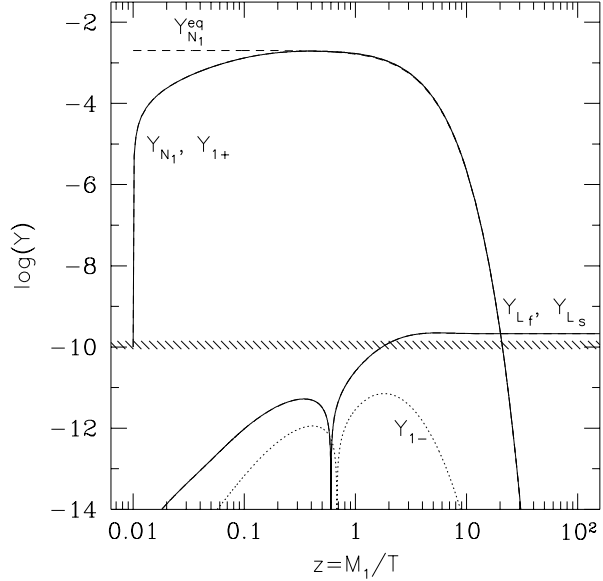,width=8.2cm}
     \end{center}
     \caption{\it Time evolution of the neutrino and the scalar neutrino
       number densities, and of the lepton asymmetries for $\l\simeq 0.1$
       and $m_3\simeq m_t$. \label{asyS}}
  \end{figure}
  
The large mass $M_3$ of the heavy Majorana neutrino $N_3$
(cf.~(\ref{nmass})), suggests that $B-L$ is already broken at the
unification scale $\Lambda_{\mbox{\scriptsize GUT}} \sim 10^{16}$
GeV, without any intermediate scale of symmetry breaking. This large
value of $M_3$ is a consequence of the choice $m_3 \simeq m_t$. This
is indeed necessary in order to obtain  sufficiently large CP asymmetry.

The recently reported atmospheric neutrino anomaly\cite{kamio} may be
due to $\n_\m$-$\n_\t$ oscillations. The required mass difference is
$\Delta m^2_{\n_\m \n_\t} \simeq (5\cdot 10^{-4}-6\cdot 10^{-3})$ eV$^2$,
together with a large mixing angle $\sin^2{2\Theta}>0.82$. In the case
of hierarchical neutrinos this corresponds to a $\t$-neutrino mass 
$m_{\n_\t} \sim (0.02-0.08)$ eV. Within the theoretical uncertainties
this is consistent with the $\t$-neutrino mass (\ref{nmass}) obtained from 
baryogenesis. The $\n_\t$-$\n_\m$ mixing angle is not constrained by 
leptogenesis and therefore a free parameter in principle. The large value,
however, is different from the mixing angles known in the quark sector
and requires an explanation. An possible framework are U(1) family
symmetries\cite{sat}. A large mixing angle can also naturally occur 
together with a mass hierarchy of light and heavy Majorana 
neutrinos\cite{kug,buy}.

Without an intermediate scale of symmetry breaking, the unification
of gauge couplings appears to require low-energy supersymmetry.
Supersymmetric leptogenesis\cite{camp} has recently been studied in detail,
taking all relevant scattering processes into account, which is necessary 
in order to get a reliable relation between the input parameters and the 
final baryon asymmetry\cite{plue}. It turns out that the lepton number 
violating scatterings are qualitatively more important than in the 
non-supersymmetric case and that they can also account for the generation of 
an equilibrium distribution of heavy neutrinos at high temperatures.
   
The supersymmetric generalization of the lagrangian (\ref{yuk}) is
the superpotential
\beq
    W = {1\over2}N^cMN^c + \m H_1\e H_2 + H_1 \e L \l_l E^c 
        + H_2 \e L \l_{\n} N^c\;,
\eeq
where, in the usual notation, $H_1$, $H_2$, $L$, $E^c$ and $N^c$ are
chiral superfields describing spin-0 and spin-${1\over 2}$ fields.
The vacuum expectation value $v_2=\left\langle H_2\right\rangle$ of the 
Higgs field $H_2$ generates the Dirac mass matrix $m_D=\l_{\n}v_2$ for 
the neutrinos and their scalar partners.

The heavy neutrinos $N_i$ and their scalar partners $\wt{N_i}$ decay with
different probabilities into final states with different lepton number. 
The generated lepton asymmetries are shown in fig.~\ref{asyS}\cite{plue}. 
$Y_{L_f}$ and $Y_{L_s}$ denote the absolute values of the asymmetries stored 
in leptons and their scalar partners, respectively. They are related
to the baryon asymmetry by
\beq
    Y_B = - {8\over 23}\ Y_L\quad, \qquad Y_L = Y_{L_f} + Y_{L_s}\;.
\eeq 
$Y_{N_1}$ is the number of heavy neutrinos per comoving volume
element, and
\beq
    Y_{1\pm} = Y_{\snone}\pm Y_{\wt{N}_1},
\eeq
where $Y_{\snone}$ is the number of scalar neutrinos per comoving
volume element. As fig.~\ref{asyS} shows, the generated baryon
asymmetry has the correct order of magnitude, as in the non-supersymmetric 
case.

From the discussion of the out-of-equilibrium condition we know that the 
generated baryon asymmetry is very sensitive to the decay width $\Gamma_{D1}$ 
of $N_1$, and therfore to $\mmoo$. In fact, the asymmetry essentially depends 
on the effective neutrino mass $\wt{m}_1$\cite{plue}. For the case of 
hierarchical neutrino masses described above, one has\cite{bp}
\beq
\wt{m}_1\ =\ {\mmoo\over M_1}\ \simeq\ m_{\n_\m}\;.
\eeq
It turns out that a sufficiently large baryon asymmetry is generated in the
range\cite{plue}
\beq
    10^{-5}\;\mbox{eV}\;\ltap\;\wt{m}_1\;\ltap\;
    5\cdot10^{-3}\;\mbox{eV}\;,
\eeq
which is consistent with the rough bound (\ref{ooeb}).
This result is independent of any assumptions on the mass matrices. It is 
very interesting that the $\n_{\m}$-mass preferred by the MSW explanation 
of the solar neutrino deficit lies indeed in the interval allowed by 
baryogenesis!

Comparing non-supersymmetric and supersymmetric leptogenesis one sees 
that the larger $CP$ asymmetry and the additional contributions from the 
sneutrino decays in the supersymmetric scenario are compensated by the 
wash-out processes which are stronger than in the non-supersymmetric case. 
The final asymmetries are the same in the non-supersymmetric and in the 
supersymmetric case.

Leptogenesis can also be considered in extended models which contain
heavy SU(2)-triplet Higgs fields in addition to right-handed 
neutrinos\cite{ma,laz}. Decays of the heavy scalar bosons can in principle
also contribute to the baryon asymmetry. However, since these Higgs particles
carry gauge quantum numbers they are strongly coupled to the plasma and
it is difficult to satisfy the out-of-equilibrium condition. The resulting
large baryogenesis temperature is in conflict with the 
`gravitino constraint'\cite{del}.

\section{SUSY mass spectrum and dark matter}

The out-of-equilibrium condition for the decay of the heavy Majorana
neutrinos, the see-saw mechanism and the experimental evidence for small
neutrino masses are all consistent and suggest rather large heavy neutrino
masses and a correspondingly large baryogenesis temperature. Within the
ansatz described in the previous section one obtains
\beq
T_B\ \sim\ M_1\sim\ 10^{10}\ \mbox{GeV}\;.
\eeq
Such a large baryogenesis temperature can only be avoided in the very
special case of a strong resonant amplification of the CP violating 
decays\cite{pil}. 

In the particularly attractive supersymmetric version of leptogenesis one 
also has to consider the following two issues: the size of other
possible contributions to the baryon asymmetry and the consistency of
the large baryogenesis temprature with the `gravitino constraint'\cite{khl}.
A large asymmetry may potentially be generated by coherent oscillations
of scalar fields which carry baryon and lepton number\cite{din}. However,
it appears likely that
the interactions of the right-handed neutrinos are sufficient to erase
such large primordial baryon and lepton asymmetries\cite{jak}. 

\begin{figure}[t]
\begin{center}
\epsfig{file=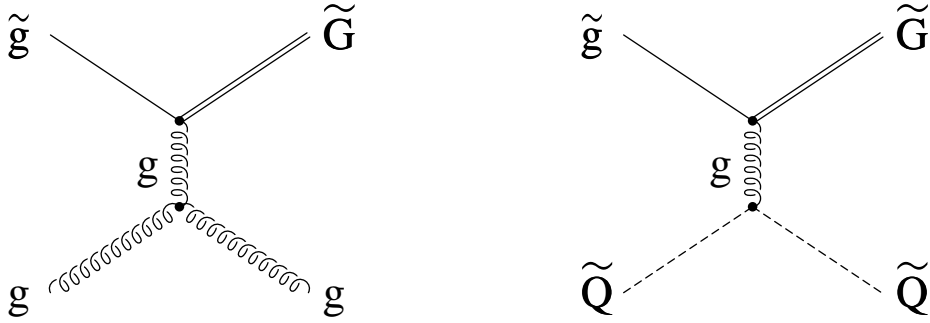}
\end{center}
\caption{\it Typical gravitino production processes mediated by gluon 
exchange. \label{gravprod}}
\end{figure}

The `gravitino contraint' is particularly interesting since it is model 
independent to a large extent, and it therefore provides very interesting
information about possible extensions of the standard model.
The production of gravitinos ($\gr$) at high temperatures is dominated
by two-body processes involving gluinos ($\gl$) (cf.~fig.~\ref{gravprod}). 
On dimensional grounds the production rate has the form
\beq
\G(T)\ \propto\ {1\over M^2}\ T^3\;,
\eeq
where $M = (8\pi G_N)^{-1/2} = 2.4\cdot 10^{18}$ GeV is the Planck mass. 
Hence, the density of thermally produced gravitinos increases strongly with 
temperature.

The production cross section is enhanced by a factor 
$(m_{\gl}/m_{\gr})^2$ for light gravitinos\cite{mor}.
The thermally averaged cross section has recently been evaluated for arbitrary
gravitino masses. The result reads\cite{bol}
\beqa
C(T) &=& \VEV{\s_{(L)} v\sbs{rel}} \NO\\
&=& {21 g^2(T)\over 32\pi\z^2(3) M^2}((N^2-1)C_A + 2n_f N C_F)
\left(1+{m_{\gl}^2(T) \over 3 m_{\gr}^2}\right)\NO\\
&&\hspace{2cm}\left(\ln{1\over g^2(T)} + {5\over 2} + 2\ln{2} -2\g_E\right)\;.
\eeqa
Here $C_A$ and $C_F$ are the usual colour factors for the group SU(N) and 
$2n_f$ is the number of colour-triplet chiral multiplets, i.e. $2n_f=12$ in 
the MSSM. The logarithmic collinear singularity of the cross section has
been regularized by a plasma mass $m \sim g(T)T$ for the gluon. 
The unknown constant part of the thermally averaged cross section is expected
to be of the same size as the term proportional to $\ln(1/g^2(T))$.
For QCD (N=3) one has
\beq\label{eq:2}
C(T) \simeq 10 {g^2(T)\over M^2}\left(1+{m_{\gl}^2(T) \over 3 m_{\gr}^2}\right)
\left(\ln{1\over g^2(T)} + 2.7\right)\;.
\eeq.

The cross section $C(T)$ enters in the Boltzmann equation, which 
describes the generation of a gravitino density $n_{\gr}$ in the thermal 
bath\cite{kol},
  \beq
    \frac{dn_{\gr}}{dt} + 3 H n_{\gr} = C(T) n\sbs{rad}^2\;.
    \label{eq:3}
  \eeq
Here $H(T)$ is the Hubble parameter and $n\sbs{rad}=\frac{\z(3)}{\p^2}T^3$ 
is the number density of a relativistic bosonic degree of freedom. 
From eqs.~(\ref{eq:2}) and (\ref{eq:3}) one obtains for the density of
light gravitinos and the corresponding contribution to $\O h^2$ at
temperatures $T<1$~MeV, i.e. after nucleosynthesis, 
  \beq
    Y_{\gr}\simeq 3.2\cdot 10^{-10}
    \left(\frac{T_B}{10^{10}\,\mbox{GeV}}\right)
    \left(\frac{100\,\mbox{GeV}}{m_{\gr}}\right)^2
    \left(\frac{m_{\gl}(\m)}{1\,\mbox{TeV}}\right)^2,
    \label{eq:6}
  \eeq

  \beqa
    \O_{\gr}h^2 & = & m_{\gr} Y_{\gr}(T) n\sbs{rad}(T) \r_c^{-1} \NO \\
    & \simeq & 0.60
    \left(\frac{T_B}{10^{10}\,\mbox{GeV}}\right)
    \left(\frac{100\,\mbox{GeV}}{m_{\gr}}\right)
    \left(\frac{m_{\gl}(\m)}{1\,\mbox{TeV}}\right)^2.    
    \label{eq:7}
  \eeqa
Here we have used $g(T_B)=0.85$; $\r_c=3H_0^2M^2$ is the critical energy
density, and $m_{\gl}(T)=\frac{g^2(T)}{g^2(\m)} m_{\gl}(\m)\gg m_{\gr}$,
with $\m \sim 100$GeV. 

The primordial synthesis of light elements (BBN) yields stringent
constraints on the amount of energy which may be released after
nucleosynthesis by the decay of heavy nonrelativistic particles into
electromagnetically  and strongly interacting relativistic particles. These 
constraints have been studied in detail by several groups\cite{ell,kaw,holt}. 
Depending on the lifetime of the decaying particle $X$ its
energy density cannot exceed an upper bound. Sufficient conditions\cite{ell}
are
  \beqa
    \mbox{(I)} & m_X Y_X(T) < 4\cdot 10^{-10}\,\mbox{GeV}, &
    \t < 2\cdot 10^6\,\mbox{sec},
    \label{eq:8} \\
    \mbox{(II)} & m_X Y_X(T) < 4\cdot 10^{-12}\,\mbox{GeV}, &
    \t\,\,\mbox{arbitrary},
    \label{eq:9}
  \eeqa
where $Y_X(T) = n_X(T)/n\sbs{rad}(T)$.

\begin{figure}[tb]
\begin{center}
    \vskip .1truein
    \centerline{\epsfysize=10cm {\epsffile{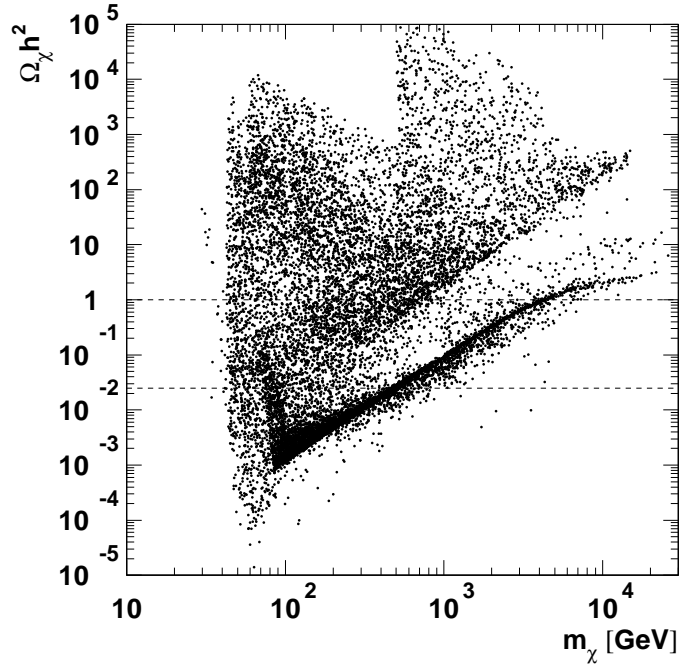}}}
    \vskip -0.1truein
    \caption{\it Neutralino relic density versus neutralino mass. The
     horizontal lines bound the region $0.025<\O_{\chi}h^2<1$.}\label{vary}
\end{center}
\vspace{-0.5cm}
\end{figure}

Gravitinos interact only gravitationally. Hence, their existence leads
almost unavoidably to a density of heavy particles which decay after
nucleosynthesis. The partial width for the decay of an unstable
gravitino into a gauge boson $B$ and a bino $\bi$ is given by
($m_{\bi}\ll m_{\gr}$)\cite{kaw},
  \beq
    \G(\gr \rightarrow B\bi)  \simeq  
    \frac{1}{32\pi}\frac{m_{\gr}^3}{M^2} 
    \simeq  \left[ 4\cdot10^8\left( \frac{100\,\mbox{GeV}}{m_{\gr}}
    \right)^3\,\mbox{sec}\right]^{-1}.
    \label{eq:10}
  \eeq
If for a fermion $\j$ the decay into a final state with a scalar $\f$ in
the same chiral multiplet and a gravitino is kinematically allowed, 
the partial width reads ($m_{\j}\gg m_{\f}$),
\beq
    \G(\j \rightarrow \gr \f) =\G(\j \rightarrow \gr \f^*)
     \simeq \frac{1}{96\pi}\frac{m_{\j}^5}{m_{\gr}^2M^2}\;. 
    \label{eq:11}
  \eeq
Given these lifetimes and the mass spectrum of superparticles in the
MSSM one can examine whether one of the conditions (I) and (II)
on the energy density after nucleosynthesis is satisfied.

Consider first a typical example of supersymmetry breaking masses in
the MSSM, $m_{\bi}<m_{\gr}\simeq 100\,\mbox{GeV}<m_{\gl}\simeq 500\,
\mbox{GeV}$, and $T_B \simeq 10^{10}$~GeV.
From eqs.~(\ref{eq:6}) and (\ref{eq:10}) we conclude $\t_{\gr}\simeq
4\cdot10^8$~sec, $m_{\gr}Y_{\gr}(T)\simeq 4\cdot10^{-9}$~GeV. 
According to condition (II) (\ref{eq:9}) this energy density
exceeds the allowed maximal energy density by 3 orders of
magnitude. This clearly illustrates the `gravitino problem'!
The stringent constraints from BBN can be evaded for very light 
gravitinos\cite{pag}, with $m_{\gr}< 1$ keV and possibly also for very heavy
gravitinos, with $m_{\gr} = {\cal O}(10$~TeV). 

\begin{figure}[tb]
\begin{center}
    \vskip .1truein
    \centerline{\epsfysize=10cm {\epsffile{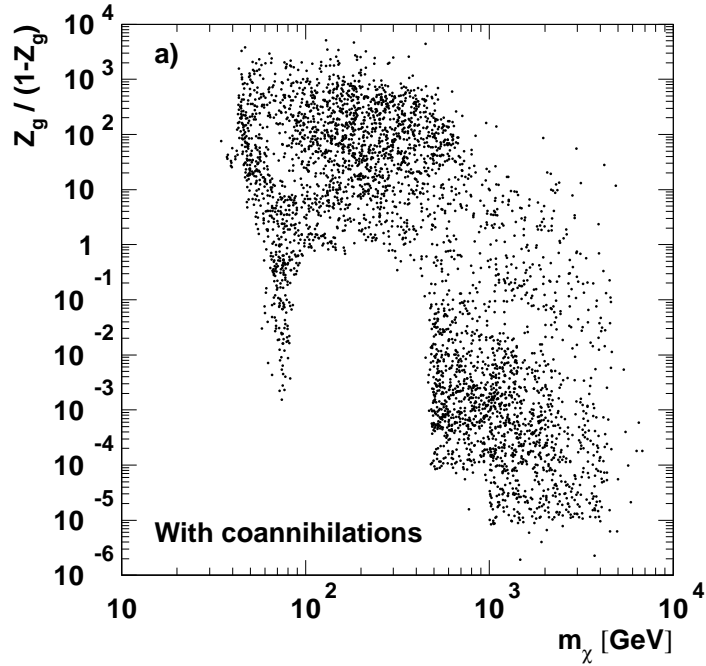}}}
    \vskip -0.1truein
    \caption{\it Neutralino composition $Z_g/(1-Z_g)$ versus neutralino mass
     for $0.025<\O_{\chi}h^2<1$.}\label{hole}
\end{center}
\vspace{-0.5cm}
\end{figure}

\begin{figure}[tb]
\begin{center}
   \vskip .1truein
    \centerline{\epsfysize=10cm {\epsffile{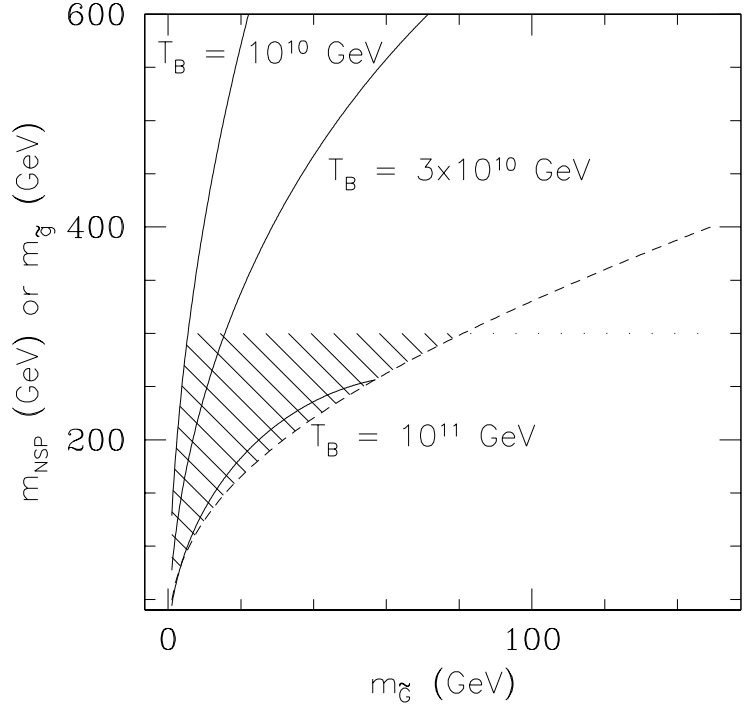}}}
    \vskip -0.3truein
\caption{\it Upper and lower bounds on the NSP mass as function of the
gravitino mass. The full lines represent the upper bound on the gluino 
mass $m_{\gl}(\m) > m_{NSP}$ for different reheating 
temperatures. The dashed line is the lower bound on $m_{NSP}$ which follows
from the NSP lifetime. A higgsino-like NSP with a mass in the shaded area
satisfies all cosmological constraints including those from primordial
nucleosynthesis.}
    \label{fig:1}
\end{center}  
\vspace{-0.5cm}
\end{figure}

Another interesting possiblity is the case where the gravitino is the LSP
with a mass ${\cal O}(100$~GeV)\cite{bol}. In this case
one has to worry about the decays of the next-to-lightest
superparticle (NSP) after nucleosynthesis. The lifetime constraint of
condition (I), $\t\sbs{NSP}<2\cdot10^6$~sec, yields a lower bound on
the NSP mass which depends on the gravitino mass $m_{\gr}$ 
(cf.~eq.~(\ref{eq:11}) and fig.~\ref{fig:1}).

For a large range  of parameters the NSP is a neutralino $\c$, 
i.e. a linear combination of higgsinos and gauginos,
  \beq
    \c = N_1 \bi + N_2 \wi + N_3 \hi + N_4 \hj\;. 
    \label{eq:12}
  \eeq
The NSP density after nucleosynthesis has been studied in great detail
by a number of authors \cite{dre}, since the density of stable neutralinos
would contribute to dark matter. A systematic study for a large range of
MSSM parameters has been carried out by Edsj\"o and Gondolo\cite{eds}. 
Varying the MSSM parameters, $\O_\c h^2$ varies over eight orders of 
magnitude, from $10^{-4}$ to $10^4$ (see fig.~\ref{vary}\cite{eds}). Note, 
that for a large part of parameter space one has $\O_\c h^2 < 0.025$. In 
particular this is the case for a higgsino-like neutralino, i.e. 
$Z_g=|N_1|^2+|N_2|^2<\frac{1}{2}$, in the mass range
$80\,\mbox{GeV}<m_\c <450\,\mbox{GeV}$\cite{eds}. For these parameters 
neutralino pair annihilation into $W$ boson pairs is very efficient, which
leads to the `higgsino hole' in fig.~\ref{hole}\cite{eds}.

The bound $\O h^2<0.008$, which corresponds to the bound on the mass density 
$m_\c Y_\c (T)<4\cdot10^{-10}$~GeV of condition (I),
is satisfied for higgsino-like neutralinos
in the mass range $80\,\mbox{GeV}<m_\c <300\,\mbox{GeV}$\cite{gon}.
We conclude that higgsino-like NSPs in this mass range and with a lifetime
$\t<2\cdot10^6$~sec are compatible with the constraints from
primordial nucleosynthesis.
Note that this is a sufficient, yet not necessary condition for satisfying
the bound $\O h^2<0.008$. Very small neutralino densities are also obtained
for other sets of MSSM parameters.  
  \begin{figure}[tb]
\begin{center}
    \vskip .1truein
    \centerline{\epsfysize=10cm {\epsffile{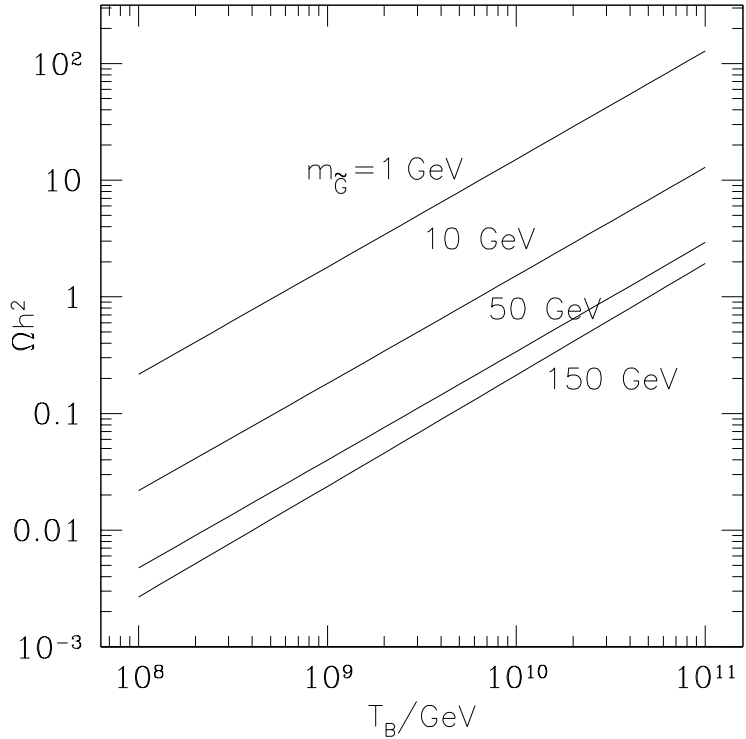}}}
    \vskip -0.1truein
    \caption{\it Contribution of gravitinos to the density parameter $\O h^2$
     for different gravitino masses $m_{\gr}$ as function of the reheating
     temperature $T_B$. The gluino mass has been set to $m_{\gl}(\m)=500$~GeV.}
    \label{fig:2}
\end{center}
\vspace{-0.5cm}
  \end{figure}

Finally, we have to discuss the necessary condition that gravitinos do
not overclose the universe,
  \beq
    \O_{\gr} h^2 < 1\;.
    \label{eq:13}
  \eeq
Since $m_{\gr}<m_\c <m_{\gl}$, the gravitino density is given by
eq.~(\ref{eq:7}). Hence, the condition (\ref{eq:13}) yields an upper bound on
the NSP mass $m_\c$ which depends on the gravitino mass and the baryogenesis 
temperature. The different constraints are summarized in 
fig.~\ref{fig:1}\cite{bol} which illustrates that for a wide range of MSSM 
parameters, where
  \beq
    m_{\gr}<m_\c <m_{\gl}\;,  \label{eq:14}
  \eeq
and $80\,\mbox{GeV}<m_\c <300\,\mbox{GeV}$, the baryogenesis
temperature may be as large as $\co(10^{10})$~GeV.

It is remarkable that for temperatures $T_B=10^8\dots 10^{11}$~GeV,
which are natural for leptogenesis, and for gravitino masses in the range
$m_{\gr}=10^0\dots 10^3$~GeV, which is expected for gravity induced
supersymmetry breaking, the relic density of gravitinos is cosmologically
important (cf.~fig.~\ref{fig:2}).
As an example, for $T_B\simeq10^{10}$~GeV, $m_{\gl}(\m)\simeq 500$~GeV, and
$m_{\gr}\simeq 50$~GeV, one has $\O_{\gr} h^2 \simeq 0.30$.\\

\section{Summary}

Detailed studies of the thermodynamics of the electroweak interactions at
high temperatures have shown that in the standard model and most of its
extensions the electroweak transition is too weak to affect the 
cosmological baryon asymmetry. Hence, one has to search for baryogenesis
mechanisms above the Fermi scale. 

Due to sphaleron processes baryon number and lepton number are related
in the high-temperature, symmetric phase of the standard model. As a
consequence, the cosmological baryon asymmetry is related to neutrino
properties. Baryogenesis requires lepton number violation, which occurs
in extensions of the standard model with right-handed neutrinos and
Majorana neutrino masses. 

Although lepton number violation is needed in order to obtain a baryon
asymmetry, it must not be too strong since otherwise any baryon and lepton
asymmetry would be washed out. This leads to stringent upper bounds on
neutrino masses which depend on the particle content of the theory.

The solar and atmospheric neutrino deficits can be interpreted as a result
of neutrino oscillations. For hierarchical neutrinos the corresponding
neutrino masses are very small. Assuming the see-saw mechanism, this suggests
the existence of very heavy right-handed neutrinos and a large scale of
$B-L$ breaking.

It is remarkable that these hints on the nature of lepton number violation
fit very well together with the idea of leptogenesis.
For hierarchical neutrino masses, with $B-L$ broken at the
unification scale $\Lambda_{\mbox{\scriptsize GUT}}\sim 10^{16}\;$GeV,  
the observed baryon asymmetry $n_B/s \sim 10^{-10}$ is naturally
explained by the decay of heavy Majorana neutrinos. The corresponding
baryogenesis temperature is $T_B \sim 10^{10}$ GeV. 

In supersymmetric models implications for the mass spectrum of superparticles 
can be derived. The rather large baryogenesis temperature leads to a high
density of gravitinos. Depending on the masses of the other superparticles 
their late decay may change the primordial abundances of light elements in
disagreement with observation. This `gravitino problem' can be avoided for
very light gravitinos, with $m_{\gr}< 1$ keV, and possibly also for very
heavy gravitinos, with $m_{\gr} = {\cal O}(10$~TeV). Another interesting
possibility is that the gravitino is the LSP, with a higgsino as NSP.
It is intriguing that for a mass range $m_{\gr}=10^0\dots 10^2$~GeV and
reheating temperatures $T_B=10^8\dots 10^{11}$~GeV, which naturally occur
after inflation, one obtains a gravitino contribution to cold dark matter 
with $\O h^2 = {\cal O}(1)$.   

\mbox{ }\vspace{4ex}\\
\noindent
\setlength{\parskip}{1ex}
{\bf Acknowledgments}\\
\mbox{ }\\    
The author would like to thank H.~J.~de~Vega and N.~S\'anchez for
organizing an enjoyable and stimulating colloquium.
\mbox{ }\\\noindent

\clearpage
\mbox{ }\vspace{4ex}\\


\noindent
{\bf\Large References}
\noindent
\begin{thebibliography}{99}

\bibitem{sac}
A.~D.~Sakharov, JETP Lett.~{\bf 5} (1967) 24

\bibitem{thoo}
G.~'t~Hooft, \prl{37}{76}{8}

\bibitem{krs}
V.~A.~Kuzmin, V.~A.~Rubakov, M.~E.~Shaposhnikov,\pl{155}{85}{36}

\bibitem{arn}
P.~Arnold, D.~Son, L.~Yaffe, \pr{55}{97}{6264};\\
D.~B\"odeker, \pl{426}{98}{351};\\
J.~Ambjorn, A.~Krasnitz, \np{506}{97}{387}

\bibitem{chem}
J.~A.~Harvey, M.~S.~Turner, \pr{42}{90}{3344}

\bibitem{fy1}
M.~Fukugita, T.~Yanagida, \pr{42}{90}{1285}

\bibitem{iba}
L.~E.~Ib\'a\~nez, F.~Quevedo, \pl{283}{92}{261}

\bibitem{dol}
For a review and references, see\\
A.~D.~Dolgov, \prep{222}{92}{309};\\
V.~A.~Rubakov, M.~E.~Shaposhnikov, Phys.~Usp.~{\bf 39} (1996)461;\\
S.~J.~Huber, M.~G.~Schmidt, {\it SUSY Variants of the Electroweak Phase
Transition}, hep-ph/9809506

\bibitem{jansen} 
For a discussion and references, see \\ 
K.~Jansen, Nucl.~Phys.~B (Proc.~Supp.) 47 (1996) 196;\\
W.~Buchm\"uller, in {\it Quarks '96} (Yaroslavl, Russia, 1996) eds.
V.~A.~Matveev et al., hep-ph/9610335;\\
K.~Rummukainen, Nucl.~Phys.~B (Proc.~Suppl.) 53 (1997) 442

\bibitem{pat}
J.~C.~Pati, A.~Salam, \pr{10}{74}{275}

\bibitem{seesaw} 
T.~Yanagida, in {\it{Workshop on unified Theories}}, KEK report 
79-18 (1979) p.~95;\\
M.~Gell-Mann, P.~Ramond, R.~Slansky, in {\it{Supergravity}} (North Holland, 
Amsterdam, 1979) eds. P.~van Nieuwenhuizen, D.~Freedman, p.~315

\bibitem{fy} 
M.~Fukugita, T.~Yanagida, \pl{174}{86}{45}

\bibitem{fisch}
W.~Fischler, G.~F.~Giudice, R.~G.~Leigh, S.~Paban \pl{258}{91}{45}

\bibitem{cov}
L.~Covi, E.~Roulet, F.~Vissani, \pl{384}{96}{169};\\
M.~Flanz, E.~A.~Paschos, U.~Sarkar, \pl{345}{95}{248}; \pl{384}{96}{487} (E)

\bibitem{bp2}
W.~Buchm\"uller, M.~Pl\"umacher, \pl{431}{98}{354}

\bibitem{pil}
For a discussion and references, see\\
A.~Pilaftsis, {\it Heavy Majorana Neutrinos and Baryogenesis}, hep-ph/9812256

\bibitem{kw} 
A.~D.~Dolgov, Ya.~B.~Zeldovich, Rev.~Mod.~Phys.~{\bf 53} (1981) 1;\\
E.~W.~Kolb, S.~Wolfram, \np{172}{80}{224}; \np{195}{82}{542}(E)

\bibitem{bp} 
W.~Buchm\"uller, M.~Pl\"umacher, \pl{389}{96}{73}

\bibitem{msw}
S.~P.~Mikheyev, A.~Y.~Smirnov, \nc{9C}{86}{17};\\
L.~Wolfenstein, \pr{17}{78}{2369}

\bibitem{tot}
N.~Hata, P.~Langacker, \pr{56}{97}{6107}

\bibitem{kamio}
Super-Kamiokande Collaboration, Y.~Fukuda et al., \pr{81}{98}{62}

\bibitem{sat}
J.~Sato, T.~Yanagida, Talk at {\it Neutrino `98}, hep-ph/9809307;\\
P.~Ramond, Talk at {\it Neutrino `98}, hep-ph/9809401

\bibitem{kug}
M.~Bando, T.~Kugo, K.~Yoshioka, \prl{80}{98}{3004}

\bibitem{buy}
W.~Buchm\"uller, T.~Yanagida, DESY 98-155, hep-ph/9810308

\bibitem{camp}
B.~A.~Campbell, S.~Davidson, K.~A.~Olive, \np{399}{93}{111}

\bibitem{plue}
M.~Pl\"umacher, \np{530}{98}{207}

\bibitem{ma}
E.~Ma, U.~Sarkar, \prl{80}{98}{5716}

\bibitem{laz}
G.~Lazarides, Q.~Shafi, \pr{58}{98}{071702}

\bibitem{del}
D.~Delepine, U.~Sarkar, DESY 98-186, hep-ph/9811479

\bibitem{khl}
M.~Yu.~Khlopov, A.~D.~Linde, \pl{138}{84}{265};\\
J.~Ellis, J.~E.~Kim, D.~V.~Nanopoulos, \pl{145}{84}{181}

\bibitem{din}
I.~Affleck, M.~Dine, \np{249}{85}{361};\\
M.~Dine, L.~Randall, S.~Thomas, \np{458}{96}{291};\\
B.~A.~Campbell, M.~K.~Gaillard, H.~Murayama, K.~A.~Olive, hep-ph/9805300

\bibitem{jak}
W.~Buchm\"uller, A.~Jakov\'ac, M.~Pl\"umacher, in preparation

\bibitem{mor}
T.~Moroi, H.~Murayama, M.~Yamaguchi, \pl{303}{93}{289}

\bibitem{bol}
M.~Bolz, W.~Buchm\"uller, M.~Pl\"umacher, DESY 98-066, hep-ph/9809381 

\bibitem{kol}
E.~W.~Kolb, M.~S.~Turner, {\it The Early Universe} (Addison Wesley,
Redwood City, 1990)

\bibitem{ell}
J.~Ellis, G.~B.~Gelmini, J.~L.~Lopez, D.~V.~Nanopoulos, S.~Sarkar,
\np{373}{92}{399}

\bibitem{kaw}
M.~Kawasaki, T.~Moroi, \ptp{93}{95}{879}

\bibitem{holt}
E.~Holtmann, M.~Kawasaki, K.~Kohri, T.~Moroi, hep-ph/9805402

\bibitem{pag}
H.~Pagels, J.~R.~Primack, \prl{48}{82}{223}
  
\bibitem{dre}
For a recent review and references, see\\
M.~Drees, in {\it Particle Dark Matter Physics: An Update}, hep-ph/9804231

\bibitem{eds}
J.~Edsj\"o, P.~Gondolo, \pr{56}{97}{1879}

\bibitem{gon}
P.~Gondolo, private communication
\end{thebibliography}
\end{document}